\documentclass[preprint,showpacs,preprintnumbers,amsmath,amssymb]{revtex4}
\usepackage{graphicx}
\usepackage{epsfig}
\usepackage{amsmath}
\usepackage{ascmac} 
\usepackage{graphics}
\usepackage{ascmac}
\usepackage{color}
\usepackage{amsmath,amssymb} 
\usepackage{amssymb}
\usepackage{pifont}
\newcommand{\cmark}{\ding{51}}%
\newcommand{\xmark}{\ding{53}}%

\def\lesssim{\ \raise.3ex\hbox{$<$}\kern-0.8em\lower.7ex\hbox{$\sim$}\ }
\def\gesim{\ \raise.3ex\hbox{$>$}\kern-0.8em\lower.7ex\hbox{$\sim$}\ }

\begin{document}
\title{Comparative studies of many-body corrections to an interacting Bose condensate}
\author{Shohei Watabe$^{1}$} 
\author{Yoji Ohashi$^{2}$}
\affiliation{$^1$ Department of Physics, The University of Tokyo, Tokyo 113-0033, Japan,
\\
$^2$ Faculty of Science and Technology, Keio University, 3-14-1 Hiyoshi, Kohoku-ku, Yokohama 223-8522, Japan.}

\begin{abstract} 
We compare many-body theories describing fluctuation corrections to the mean-field theory in a weakly interacting Bose-condensed gas. Using a generalized random-phase approximation, we include both density fluctuations and fluctuations in the particle-particle scattering channel in a consistent manner. We also separately examine effects of the fluctuations within the framework of the random-phase approximation. Effects of fluctuations in the particle-particle scattering channel are also separately examined by using the many-body $T$-matrix approximation. We assess these approximations with respect to the transition temperature $T_{\rm c}$, the order of phase transition, as well as the so-called Nepomnyashchii-Nepomnyashchii identity, which states the vanishing off-diagonal self-energy in the low-energy and low-momentum limit. Since the construction of a consistent  theory for interacting bosons which satisfies various required conditions is a long standing problem in cold atom physics, our results would be useful for this important challenge.  
\end{abstract}

\maketitle
\section{Introduction}
Since the realization of the Bose-Einstein condensation (BEC) in ultracold atomic gases, various interesting phenomena observed in this system have been theoretically explained. Even now, however, the theory of a weakly interacting Bose gas still has room for improvement. Indeed, while the mean-field Bardeen-Cooper-Schrieffer (BCS) theory is applicable to the entire temperature region below the transition temperature $T_{\rm c}$, the applicability of mean-field theories for a condensed Bose gas is restricted to the region far below $T_{\rm c}$, because they unphysically give the first order phase transition~\cite{Shi1998,Reatto1969}. In addition, the Hartree--Fock--Bogoliubov (HFB) theory does not satisfy the Hugenholtz-Pines theorem~\cite{Hugenholtz1959}, leading to gapped excitations below $T_{\rm c}$. This unphysical result is absent in the Bogoliubov approximation~\cite{Bogoliubov1947} and the (HFB)-Popov approximation~\cite{Popov1964A,Popov1964B,Popov1983BOOK,Griffin1996,Shi1998} (i.e., Shohno model~\cite{Shohno1964}). 
However, they do not satisfy the identity proved by Nepomnyashchii and Nepomnyashchii~\cite{Nepomnyashchii1975,Nepomnyashchii1978}, stating the vanishing off-diagonal self-energy in the low-energy and low-momentum limit in the BEC phase. 
\par
Another problem associated with these mean-field theories is that they cannot correctly describe interaction corrections to $T_{\rm c}$, although more sophisticated approaches predict the deviation of $T_{\rm c}$ from an ideal Bose gas result ($\equiv T_{\rm c}^{0}$). In a trapped gas, $T_{\rm c}$ is lowered, because the density profile spreads out by a repulsive interaction, leading to the decrease in the central particle density~\cite{Grossmann1995,Giorgini1996,Ketterle1996,Haugerud1997}. In the uniform system, the enhancement of $T_{\rm c}$ has been predicted in the region of small gas parameter~\cite{Stoof1992,Gruter1997,Baym1999,Holzmann1999,Arnold2000,Arnold2001,Baym2001,Kashurnikov2001,Kneur2002,Davis2003,Kleinert2003,Andersen2004,Kastening2004,Nho2004,Ledowski2004,Blaizot2005,Blaizot2011,Blaizot2012,Tsutsui2012}. In particular, Monte-Carlo simulations give $(T_{\rm c}-T_{\rm c}^{0})/ T_{\rm c}^{0} = c_{1} an^{1/3}$ with $c_{1} \simeq 1.3$~\cite{Arnold2001,Kashurnikov2001,Nho2004}. In the strongly correlated case (which corresponds to liquid $^4$He), however, it has been pointed out that $T_{\rm c}$ is suppressed by mass enhancement~\cite{Gruter1997,Fyenman1953}. 
\par
Toward the construction of a consistent theory of a weakly-interacting Bose gas, we investigate many-body effects on this system. In this regard, we recall that, within the framework of the many-body $T$-matrix approximation, Shi and Griffin showed the vanishing effective interaction in the low-energy and static limit at $T_{\rm c}$~\cite{Shi1998}, which is never obtained in the mean-field theories mentioned above. Using this, they obtained the expected second-order phase transition, although their many-body $T$-matrix theory with the static approximation still gives the same value of $T_{\rm c}$ as that in an ideal Bose gas. 
\par
Stimulated by this many-body approach~\cite{Shi1998}, in this paper, we extend it to include fluctuation corrections beyond the static approximation. We include fluctuations in both the particle-particle scattering channel and density channel within the framework of the generalized random phase approximation (GRPA). To examine effects of fluctuations in each channel, we also consider the case with the former fluctuations by using the many-body $T$-matrix approximation (MBTA). Effects of the latter fluctuations are also separately examined within the framework of the random phase approximations (RPA). 
\par
In order to achieve them, we develop the 4$\times$4 matrix formalism, which includes 
all the possible polarization functions in four-point vertex functions. 
By using this formalism, we develop the GRPA and extend the MBTA as well as the RPA. 
These diagrammatic contributions are first presented in this paper. In fact, this matrix formalism cannot be seen previously in the field of the BEC, to the best of our knowledge. 
\par 
Treating these many-body theories without employing the static approximation, we evaluate $T_{\rm c}$, as well as the order of phase transition. 
The critical temperature shift has been evaluated at most in the static limit (\cite{Andersen2004} and the references therein). The order of the phase transition has been discussed also in the static limit based on the many-body theory above $T_{\rm c}$~\cite{Tsutsui2012}. We assess these problems using our formalism below $T_{\rm c}$. 
We also examine if they satisfy the Nepomnyashchii-Nepomnyashchii identity, stating the vanishing off-diagonal self-energy in the low-energy and low-momentum limit. 
\par 
In contrast to the earlier study of the MBTA~\cite{Shi1998}, the critical temperature in our calculation is shifted, 
because we extend this formalism beyond the static approximation. 
In this earlier study~\cite{Shi1998}, the infrared divergence in the polarization function was omitted by hand to obtain the non-vanishing off-diagonal self-energy. In this study, we do not apply this {\it ad hoc} omission. 
\par 
In section II, we explain the many-body theories used in this paper. 
We show our numerical results for $T_{\rm c}$ in section III. 
The condensate fraction below $T_{\rm c}$ is discussed in section IV. 
In section V, we assess the many-body theories on the viewpoint of the Nepomnyashchii-Nepomnyashchii identity. 
In this paper, we set $\hbar = k_{\rm B} = 1$, and the system volume $V$ is taken to unity. 
\par
\section{Formulation}
\par
We consider a weakly-interacting Bose gas, described by the Hamiltonian,
\begin{equation}
H=\sum_{\bf p} (\varepsilon_{\bf p} - \mu)
a_{\bf p}^{\dag} a_{\bf p} + \frac{U}{2} 
\sum_{{\bf p},{\bf p}',{\bf q}} 
a_{{\bf p} + {\bf q}}^{\dag} a_{{\bf p}'-{\bf q}}^{\dag} a_{{\bf p}'} a_{{\bf p}}. 
\label{eq1}
\end{equation} 
Here, $a_{\bf p}^\dagger$ is the creation operator of a Bose atom with the kinetic energy $\varepsilon_{\bf p}-\mu = {\bf p}^{2}/(2m)-\mu $, measured from the chemical potential $\mu$ (where $m$ is an atomic mass). $U$ is a repulsive interaction, which is related to the $s$-wave scattering length $a$ as
\begin{align} 
\frac{4 \pi a}{ m}
= \frac{ U}{ 1+ \displaystyle{ U }
 \sum_{{\bf p}}^{p_{\rm c}} 
 \displaystyle{ \frac{1}{2 \varepsilon_{\bf p}} }
 }, 
\label{eq3}
\end{align} 
where $p_{\rm c}$ is a cutoff. 
\par
As usual, the order parameter (which is also referred to as the condensation fraction $n_0$) is introduced by the Bogoliubov prescription~\cite{Bogoliubov1947}. That is, $a_{{\bf p} = {\bf 0}}$ and $a_{{\bf p} = {\bf 0}}^{\dag}$ are replaced by $\sqrt{n_{0}}$. Physical properties in the BEC phase is conveniently described by the $2\times 2$-matrix single-particle thermal Green's function,
\begin{equation}
G (p) = \frac{1}{i \omega_{n} \sigma_{3} - \varepsilon_{\bf p} + \mu - \Sigma (p)}. 
\label{Eq37}
\end{equation} 
Here, we have simply written $p = ({\bf p}, i \omega_{n})$, where $\omega_{n}$ is the boson Matsubara frequency. $\sigma_j$ ($j=1,2,3$) are the Pauli matrices, acting on the space spanned by $(a_{\bf p},a_{-{\bf p}}^\dagger)$. The $2\times 2$-matrix self-energy satisfies $\Sigma_{22}(p)=\Sigma_{11}(-p)$ and $\Sigma_{21}(p)=\Sigma_{12}(-p)$~\cite{Shi1998}. The diagonal component of Eq. (\ref{Eq37}) is related to the non-condensate density $n'=n-n_0$ as (where $n$ is the total particle density)
\begin{equation}
n' = - T \sum_{p} G_{11} (p) e^{i\omega_{n} \delta}. 
\end{equation} 
\par
\begin{figure}
\begin{center}
\includegraphics[width=13cm]{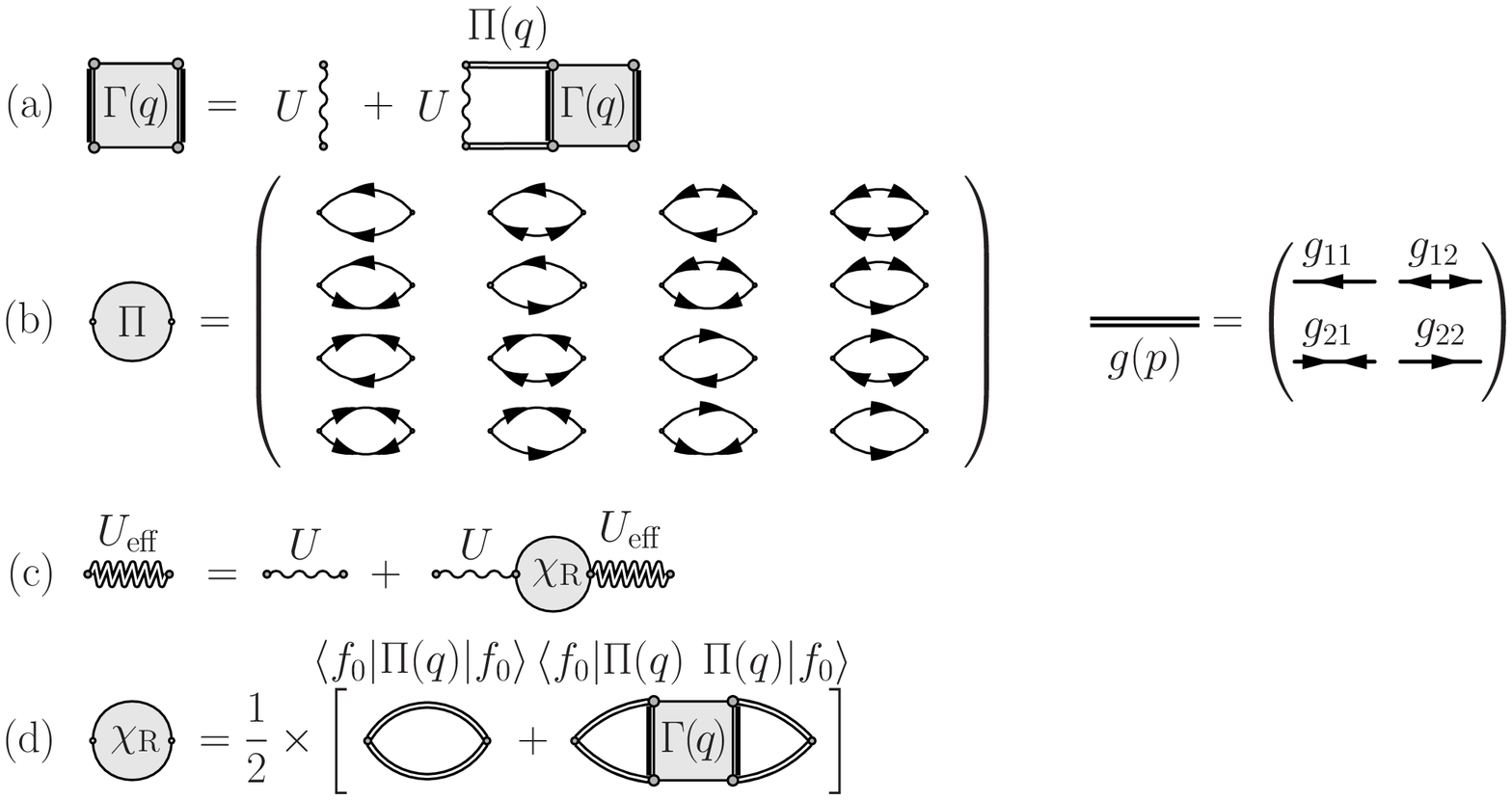}
\end{center}
\caption{
(a) Bethe-Salpeter equation of the four-point vertex function $\Gamma (q)$ in the generalize random-phase approximation (GRPA). The double solid line shows the $2\times 2$-matrix Green's function $g(p)$. The wavy line describes the repulsive interaction $U$. (b) Generalized polarization function $\Pi(q)$. Each solid line denotes a matrix element of the Green's function $g(p)$ in Eq. (\ref{EqE1}). (c) Effective interaction $U_{\rm eff} (q)$ involving density fluctuations. (d) Regular part $\chi_{\rm R}$ of the density-density correlation function~\cite{NOTE}. 
}
\label{fig1}
\end{figure} 
\par
To explain the GRPA, MBTA, and RPA, it is convenient to introduce the $4\times 4$-matrix four-point vertex $\Gamma (q)$, diagrammatically given in Fig.~\ref{fig1} (a). The explicit expression is
\begin{align}
{ \Gamma} (q) = \frac{U}{1 - U { \Pi}(q)},
\label{RPABetheSalpeter0}
\end{align} 
where  
\begin{eqnarray}
\Pi (q)=
-T\sum_p
\begin{pmatrix}
K_{1111}(p,q) & K_{1112}(p,q) & K_{1211}(p,q) & K_{1212}(p,q) 
\\
K_{1121}(p,q) & K_{1122}(p,q) & K_{1221}(p,q) & K_{1222}(p,q) 
\\
K_{2111}(p,q) & K_{2112}(p,q) & K_{2211}(p,q) & K_{2212}(p,q) 
\\
K_{2121}(p,q) & K_{2122}(p,q) & K_{2221}(p,q) & K_{2222}(p,q) 
\end{pmatrix} 
\end{eqnarray} 
is the $4\times 4$-matrix generalized polarization function diagrammatically described in Fig.~\ref{fig1} (b). 
Here, $K_{ijkl} (p,q) = g_{ij} (p+q) g_{kl} (-p)$ is the two-particle Green's function. 
$g_{ij} (p)$ is a matrix element of the one-particle Green's function in the HFB--Popov approximation, given by 
\begin{align}
g(p) = \frac{ 1 }{i \omega_{n} \sigma_{3} - \xi_{\bf p}  - U n_{0} \sigma_{1}}, \label{EqE1}
\end{align} 
where $\xi_{\bf p} = \varepsilon_{\bf p} + U n_{0}$. Using the symmetry properties $g_{22}(p) = g_{11} (-p)$ and $g_{12}(p) = g_{12}(-p)$, one finds that all the matrix elements of $\Pi$ are not independent as
\begin{align}
\Pi(q) = & 
\begin{pmatrix}
\Pi_{11} (q) & \Pi_{12} (q) & \Pi_{12} (q) & \Pi_{14} (q) \\
\Pi_{12} (q) & \Pi_{22} (q) & \Pi_{14} (q) & \Pi_{12}^{*} (q) \\
\Pi_{12} (q) & \Pi_{14} (q) & \Pi_{22} (q) & \Pi_{12}^{*} (q) \\
\Pi_{14} (q) & \Pi_{12}^{*} (q) & \Pi_{12}^{*} (q)& \Pi_{11}^{*}  (q)
\end{pmatrix}. 
\label{RPAPiMatrix}
\end{align}
(For the detailed expressions of $\Pi_{11,12,14,22}$, see appendix A.) Some of $\Gamma_{ij}(q)$ are thus also related to each other as
\begin{align}
\Gamma(q) = & 
\begin{pmatrix}
\Gamma_{11} (q) & \Gamma_{12} (q) & \Gamma_{12} (q) & \Gamma_{14} (q) \\ 
\Gamma_{12} (q) & \Gamma_{22} (q) & \Gamma_{23} (q) & \Gamma_{12}^* (q) \\ 
\Gamma_{12} (q) & \Gamma_{23} (q) & \Gamma_{22} (q) & \Gamma_{12}^* (q) \\ 
\Gamma_{14} (q) & \Gamma_{12}^{*} (q) & \Gamma_{12}^* (q) & \Gamma_{11}^{*} (q) \end{pmatrix}. 
\label{GammaMatrix}
\end{align} 
\par
We also introduce the effective interaction $U_{\rm eff} (p)$ (Fig.~\ref{fig1} (c)), describing effects of density fluctuations, given by
\begin{align}
U_{\rm eff} (p) = & \frac{U}{ 1 - U \chi_{\rm R} (p) }.
\label{RPABetheSalpeter}
\end{align} 
Here, $\chi_{\rm R}$ is the regular part of the density-density correlation function (Fig.~\ref{fig1} (d))\cite{NOTE},
\begin{align}
\chi_{\rm R} (q) = & \frac{1}{2} \langle f_0 | [ { \Pi} (q) + { \Pi} (q) { \Gamma} (q) { \Pi} (q) ] |f_0 \rangle, 
\label{eq10}
\end{align} 
where $\langle f_{0} | = (0,1,1,0)$ and $|f_0\rangle = (0, 1,1, 0)^{\rm T}$. 
\par
\begin{figure}
\begin{flushleft}
\includegraphics[width=19cm]{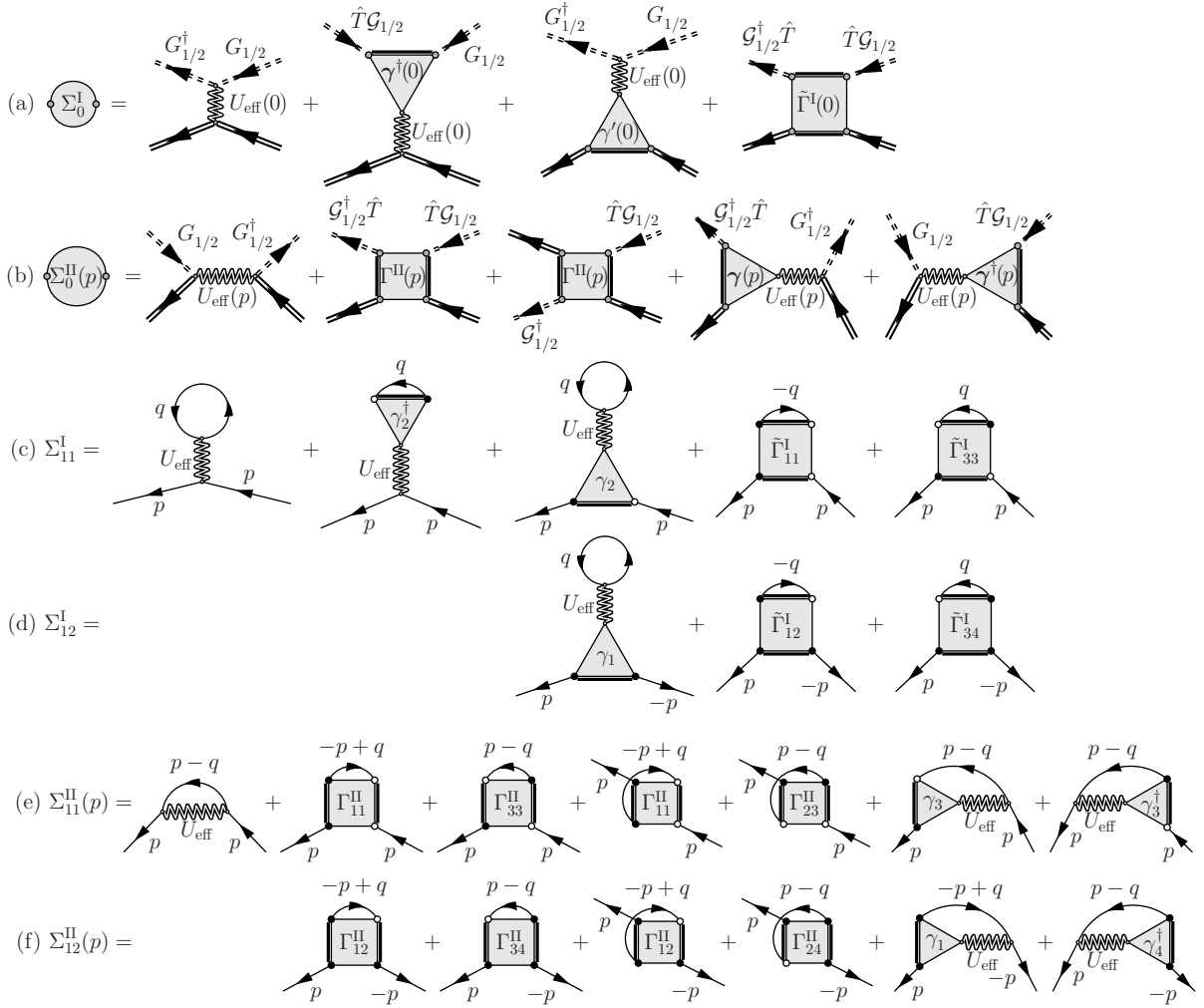} 
\end{flushleft}
\caption{Self-energy corrections in the generalized random-phase approximation (GRPA). (a) $\Sigma_{0}^{\rm I}$, (b) $\Sigma_{0}^{\rm II}$, (c) $\Sigma_{11}^{\rm I}$, 
(d) $\Sigma_{12}^{\rm I}$, (e) $\Sigma_{11}^{\rm II} (p)$, and (f) $\Sigma_{12}^{\rm II} (p)$. }
\label{fig2} 
\end{figure} 
\par
\subsection{Generalized random phase approximation (GRPA)}
\par
In the GRPA, the self-energy correction may be conveniently written as the sum of the two parts,
\begin{equation}
\Sigma(p)=\Sigma_0(p)+\Sigma'(p). 
\end{equation}
Here, $\Sigma_0(p)$ is characterized by condensate Green's functions $G_{1/2}=\sqrt{-n_0}(1,1)^{\rm T}$ and 
$G_{1/2}^{\dag} = \sqrt{-n_{0}} (1,1)^{\rm }$, and 
$\Sigma'(p)$ has an internal loop of the diagonal Green's function $g_{11}$, diagrammatically described as Figs.~\ref{fig2} (a) and (b), and (c)-(f), respectively. Each of these components can be further decomposed into the $p$-independent part $(\Sigma_{0}^{\rm I}, \Sigma^{\rm I})$ and the remaining $p$-dependent part $(\Sigma_{0}^{\rm II}, \Sigma^{\rm II})$ as
\begin{align}
\Sigma_{0} (p) = & \Sigma_{0}^{\rm I} + \Sigma_{0}^{\rm II} (p) , 
\label{eq14}
\\ 
\Sigma' (p) = & \Sigma^{\rm I} + \Sigma^{\rm II} (p). 
\label{eq15}
\end{align} 
The two components in Eq. (\ref{eq14}) are given by
\begin{align} 
{\Sigma}_{0}^{\rm I} = & 
- U_{\rm eff} (0) \frac{1}{2} { G}_{1/2}^{\dag} { G}_{1/2} 
- U_{\rm eff} (0) \frac{1}{2} \boldsymbol{\mathit\gamma}^{\dag} (0) \hat T {\mathcal G}_{1/2} { G}_{1/2} 
-  \gamma' (0) U_{\rm eff} (0)  \frac{1}{2} { G}_{1/2}^{\dag} { G}_{1/2} 
\nonumber
\\  
& 
- {\mathcal G}_{1/2}^{\dag} \hat T \tilde \Gamma^{\rm I} (0) \hat T {\mathcal G}_{1/2} , 
\label{eq1}
\\
{\Sigma}_{0}^{\rm II} (p) = &
- { G}_{1/2} U_{\rm eff }(p) { G}_{1/2}^{\dag}
- {\mathcal G}_{1/2}^{\dag} \hat T \Gamma^{\rm II} (p) \hat T {\mathcal G}_{1/2}- {\mathcal G}_{1/2}^{\dag} \Gamma^{\rm II} (p) \hat T {\mathcal G}_{1/2}
\nonumber
\\ &
- {\mathcal G}_{1/2}^{\dag} \hat T \boldsymbol{\mathit \gamma} (p) U_{\rm eff} (p) {G}_{1/2}^{\dag}
- { G}_{1/2} U_{\rm eff} (p) \boldsymbol{\mathit\gamma}^{\dag} (p) \hat T {\mathcal G}_{1/2},
\label{eq2}
\end{align}
\begin{figure}
\begin{center}
\includegraphics[width=8cm]{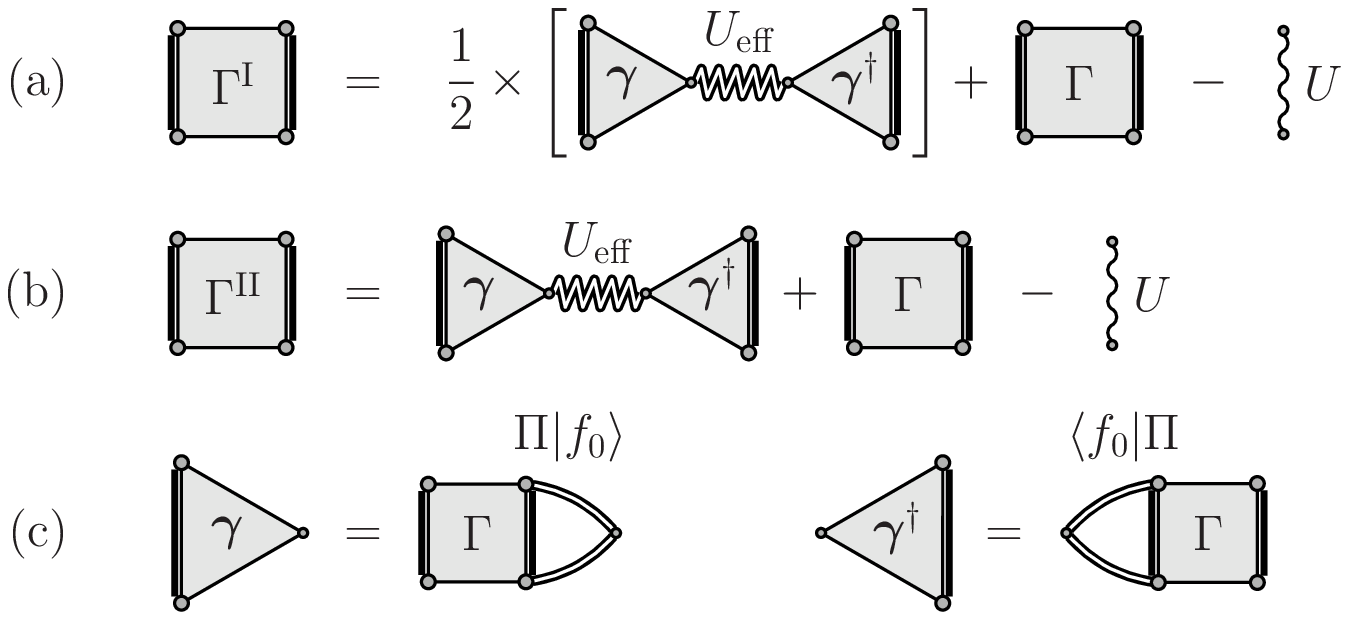}
\end{center}
\caption{Four point vertices (a) $\Gamma_{\rm I}$ and (b) $\Gamma_{\rm II}$. (c) Three point vertices $\boldsymbol{\mathit \gamma}$ and $\boldsymbol{\mathit \gamma}^{\dag}$. 
}
\label{fig3}
\end{figure} 
where
\begin{eqnarray}
{\mathcal G}_{1/2} = 
\sqrt{-n_0} 
\left(
\begin{array}{cc}
1& 0\\
1& 0\\
0& 1\\
0& 1\\
\end{array}
\right), 
\quad 
{\mathcal G}_{1/2}^{\dag} = 
\sqrt{-n_0} 
\left(
\begin{array}{cccc}
1 & 1 & 0 & 0 \\
0 & 0 & 1 & 1
\end{array}
\right),
\end{eqnarray}
\begin{align}
\hat T= 
\begin{pmatrix}
1 & 0 & 0 & 0 \\
0 & 0 & 1 & 0 \\
0 & 1 & 0 & 0 \\
0 & 0 & 0 & 1 
\end{pmatrix}. 
\end{align} 
In Eqs. (\ref{eq1}) and (\ref{eq2}), the four-point vertices $\Gamma^{\rm I} (p)$ and $\Gamma^{\rm II} (p)$ diagrammatically given by Figs.~\ref{fig3} (a) and (b), have the forms,
\begin{align}
\Gamma^{\rm I} (p) = & \frac{1}{2} \boldsymbol{\mathit\gamma} (p) U_{\rm eff} (p) \boldsymbol{\mathit\gamma}^{\dag} (p) + \Gamma (p) - U, 
\label{eq3}
\\
\Gamma^{\rm II} (p) = & \boldsymbol{\mathit\gamma} (p) U_{\rm eff} (p) \boldsymbol{\mathit\gamma}^{\dag} (p) + \Gamma (p) - U. 
\label{eq4}
\end{align}
$\tilde \Gamma^{{\rm I}} (p)$ in Eq. (\ref{eq1}) is also a four-point vertex, given by
\begin{align}
\tilde \Gamma^{{\rm I}} (p) = & 
\begin{pmatrix}
\Gamma^{\rm I}_{23} (p) & \Gamma^{\rm I}_{13} (p) & \Gamma^{\rm I}_{24} (p) & \Gamma^{\rm I}_{14} (p) \\
\Gamma^{\rm I}_{43} (p) & \Gamma^{\rm I}_{33} (p) & \Gamma^{\rm I}_{44} (p) & \Gamma^{\rm I}_{34} (p) \\
\Gamma^{\rm I}_{21} (p) & \Gamma^{\rm I}_{11} (p) & \Gamma^{\rm I}_{22} (p) & \Gamma^{\rm I}_{12} (p) \\
\Gamma^{\rm I}_{41} (p) & \Gamma^{\rm I}_{31} (p) & \Gamma^{\rm I}_{42} (p) & \Gamma^{\rm I}_{32} (p) 
\end{pmatrix} .  
\end{align} 
$\boldsymbol{\mathit \gamma}=\Gamma(q)\Pi(q)|f_0\rangle$ and $\boldsymbol{\mathit \gamma}^{\dag}=\langle f_0|\Pi(q)\Gamma(q)$ in Eqs. (\ref{eq1}), (\ref{eq2}), (\ref{eq3}) and (\ref{eq4}) are three-point vertices diagrammatically shown in Fig.~\ref{fig3} (c). $\gamma' (p)$ in Eq.(\ref{eq1}) is also a three-point vertex, having the form
\begin{align}
\gamma' (p) = & 
\begin{pmatrix}  
\gamma_{2} (p) & \gamma_{1} (p) \\
\gamma_{4} (p) & \gamma_{3} (p)
\end{pmatrix}. 
\end{align} 
\par
The self-energy corrections $\Sigma^{\rm I}$ and $\Sigma^{\rm II}$ in Eq. (\ref{eq15}) are respectively given by, 
\begin{align} 
\Sigma^{\rm I} = ( U_{11}^{\rm } \sigma_{0} + U_{12}^{\rm } \sigma_{1} ) n_{0}',
\end{align}
\begin{align}
\Sigma^{\rm II} (p) = 
\begin{pmatrix}
\Sigma_{11}^{\rm II} (p) & \Sigma_{12}^{\rm II} (p)
\\
\Sigma_{12}^{\rm II} (-p) & \Sigma_{11}^{\rm II} (-p)
\end{pmatrix}, 
\label{eq100}
\end{align}
where $n'_0=-T \sum\limits_{p} g_{11}(p) e^{i \omega_{n} \delta}$, and
\begin{align} 
U_{11}^{\rm } = & 
U_{\rm eff} (0) [1 + \gamma_{2} (0) + \gamma_{2}^{\dag} (0) ] 
+ \tilde \Gamma^{{\rm I}}_{11}(0) + \tilde \Gamma^{{\rm I}}_{33}(0),  
\nonumber
\\
U_{12}^{\rm } = & 
 U_{\rm eff} (0)  \gamma_{1} (0)  + \tilde \Gamma^{{\rm I}}_{12}(0) + \tilde \Gamma^{{\rm I}}_{34}(0).    
\end{align}
Expressions for $\Sigma_{ij}^{\rm II}(p)$ in Eq. (\ref{eq100}) are
\begin{align}
\begin{pmatrix}
\Sigma_{11}^{\rm II} (p) 
\\
\Sigma_{12}^{\rm II} (p) 
\end{pmatrix} 
= 
- T \sum\limits_{q} 
\begin{pmatrix}
A (q) & B (q) \\ C (q) & D (q) 
\end{pmatrix} 
\begin{pmatrix}
G_{11} (+p-q) 
\\
G_{11} (-p+q)
\end{pmatrix}, 
\end{align}
where 
\begin{align}
A (q) = & U_{\rm eff} (q) [1 + \gamma_{3} (q) + \gamma_{3}^{\dag} (q) ] + \Gamma^{\rm II}_{33} (p) + \Gamma^{\rm II}_{23} (p), 
\\
B (q) = & 2 \Gamma^{\rm II}_{11} (q) , 
\\
C (q) = & U_{\rm eff} (q)  \gamma_{4}^{\dag} (q)  + \Gamma^{\rm II}_{34} (q) + \Gamma^{\rm II}_{24} (q) , 
\\
D (q) = & U_{\rm eff} (q) \gamma_{1} (q) + 2 \Gamma^{\rm II}_{12} (q). 
\end{align}
\par
\subsection{Many-body $T$-matrix approximation (MBTA)}
\par
\begin{figure}
\begin{center}
\includegraphics[width=8cm]{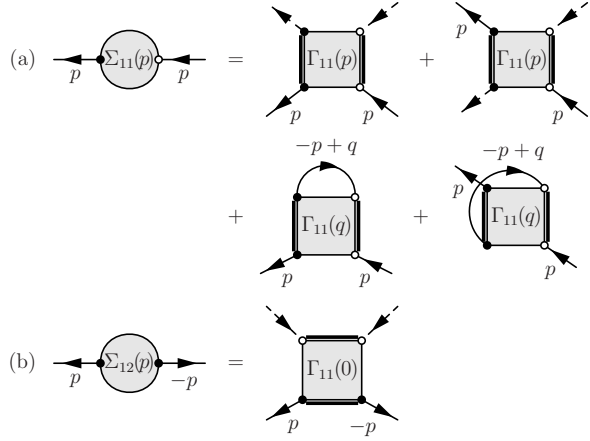}
\end{center}
\caption{Self-energy in the many-body $T$-matrix approximation (MBTA). (a) Diagonal component $\Sigma_{11}$. (b) Off-diagonal component $\Sigma_{12}$. The dashed lines describe $\sqrt{n_0}$.}
\label{fig4}
\end{figure}

The MBTA self-energy $\Sigma(p)$ involves particle-particle scattering processes~\cite{Shi1998}. Summing up the diagrams in Fig.~\ref{fig4}, we have
\begin{align}
\Sigma_{11} (p) = & 2 n_{0} \Gamma_{11} (p) 
- 2 T \sum\limits_{q} \Gamma_{11} (q) g_{11} (-p+q), 
\label{Eq243}
\\
 \Sigma_{12} (p) = & n_{0} \Gamma_{11} (0). 
\label{Eq244}
\end{align} 
In the normal state, because of $\Pi_{14} = \Pi_{12} = n_0=0$, one obtains the vanishing off-diagonal self-energy, and
\begin{align}
\Sigma_{11} (p) = - 2T 
\sum\limits_{q} 
\frac{U}{1-U \Pi_{11}(q)}g_{11} (-p+q), 
\label{Eq243b}
\end{align}
which is just the same expression as the self-energy in the ordinary many-body $T$-matrix approximation above $T_{\rm c}$.
\par
\par
The many-body $T$-matrix theory developed by Shi and Griffin~\cite{Shi1998} approximates the four-point vertex to $\Gamma_{11} = U /[1 - U \Pi_{11} (0)]$, where the static approximation is also taken. In this approximation, $\Pi_{11}(0)$ always diverges below $T_{\rm c}$, leading to the vanishing off-diagonal self-energy ($\Sigma_{12}(p)=0$). To avoid this, Ref.~\cite{Shi1998} eliminates this singularity by hand, assuming that such infrared divergence should be absent when one appropriately includes higher order corrections.
\par
In contrast to this earlier work~\cite{Shi1998}, we fully take into account the energy and momentum dependence of the particle-particle scattering vertex function $\Gamma_{11}$ in Eq. (\ref{GammaMatrix}). As shown in Appendix B, the infrared divergence in this $\Gamma_{11}$ is canceled out, so that the off-diagonal self-energy does not vanish even below $T_{\rm c}$.
\par
\par
\begin{figure}
\begin{center}
\includegraphics[width=7cm]{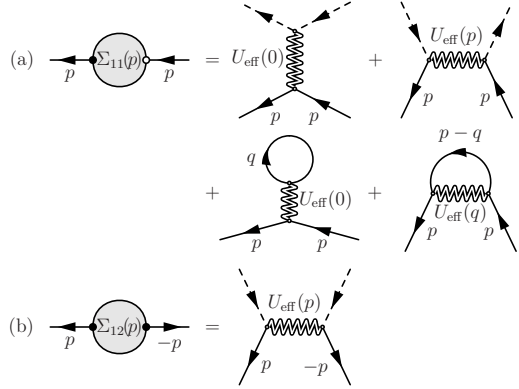}
\end{center}
\caption{Self-energy in the random phase approximation (RPA). (a) Diagonal component $\Sigma_{11}$. (b) Off-diagonal component $\Sigma_{12}$. }
\label{fig5}
\end{figure} 
\par
\subsection{Random phase approximation (RPA)}
\par
The RPA self-energy is diagrammatically given in Fig.~\ref{fig5}. Summing them up, one has
\begin{align}
\Sigma_{11} (p) = & (n_0+n'_0)U_{\rm eff} (0) + n_{0} U_{\rm eff} (p) 
- T \sum\limits_{q} U_{\rm eff} (q) g_{11} (p-q), 
\label{eq27} 
\\
\Sigma_{12} (p) = & n_{0} U_{\rm eff} (p) . 
\label{eq28} 
\end{align}  
At $T_{\rm c}$, the regular part of the density-density correlation function (\ref{eq10}) is reduced to
\begin{align} 
\chi_{\rm R} (p) = & \frac{\Pi_{22} (p) }{1-U\Pi_{22} (p) }, 
\end{align} 
which is consistent with the ordinary RPA in the normal state. When we retain the lowest-order bubble diagram in Fig.~\ref{fig1} (d), the polarization function in Eq. (\ref{RPABetheSalpeter}) is further simplified as $\chi_{\rm R} (q) = \Pi_{22} (q) $. In this paper, we also deal with this simpler version (which we call {\it the simplified RPA} (s-RPA)).  
\par 
\begin{figure}
\begin{center}
\includegraphics[width=7.5cm]{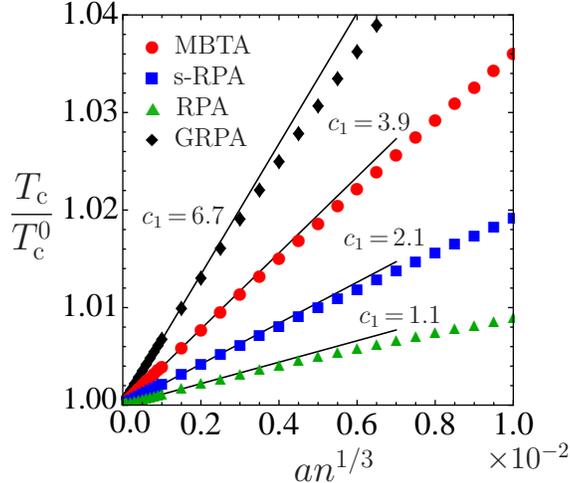}
\end{center}
\caption{(Color online) BEC phase transition temperatures $T_{\rm c}$ in a weakly-interacting Bose gas, calculated in the many-body $T$-matrix approximation (MBTA), the random-phase approximation (RPA), the simplified RPA (s-RPA), and the generalized RPA (GRPA). The interaction is measured in terms of the scattering length $a$.  In numerical calculations, we set $p_{\rm c} = 5 p_{0}$ (where $p_{0}=\sqrt{2m T_{\rm c}^{0}}$).
}
\label{fig6}
\end{figure} 
\par
\section{Phase transition temperature $T_{\rm c}$} 
\par
All the approximations (MBTA, RPA, s-RPA, and GRPA) give the enhancement of $T_{\rm c}$ by the repulsive interaction $U$ (Fig.~\ref{fig6}), which is consistent with the previous work
~\cite{Stoof1992,Gruter1997,Baym1999,Holzmann1999,Arnold2000,Arnold2001,Baym2001,Kashurnikov2001,Kneur2002,Davis2003,Kleinert2003,Andersen2004,Kastening2004,Nho2004,Ledowski2004,Blaizot2005,Blaizot2011,Blaizot2012,Tsutsui2012}. Measuring the shift of $T_{\rm c}$ from the ideal Bose gas result ($T_{\rm c}^{0}$) as 
\begin{equation}
{T_{\rm c}-T_{\rm c}^{0} \over T_{\rm c}^{0}}=c_1 an^{1/3},
\end{equation}
one finds $c_{1}=6.7$ (GRPA), 3.9 (MBTA), 2.1 (s-RPA), and 1.1 (RPA). Among them, the RPA result is closest to the Monte-Carlo result $c_1\simeq1.3$~\cite{Arnold2001,Kashurnikov2001,Nho2004}.
\par
\begin{figure}
\begin{center}
\includegraphics[width=9cm]{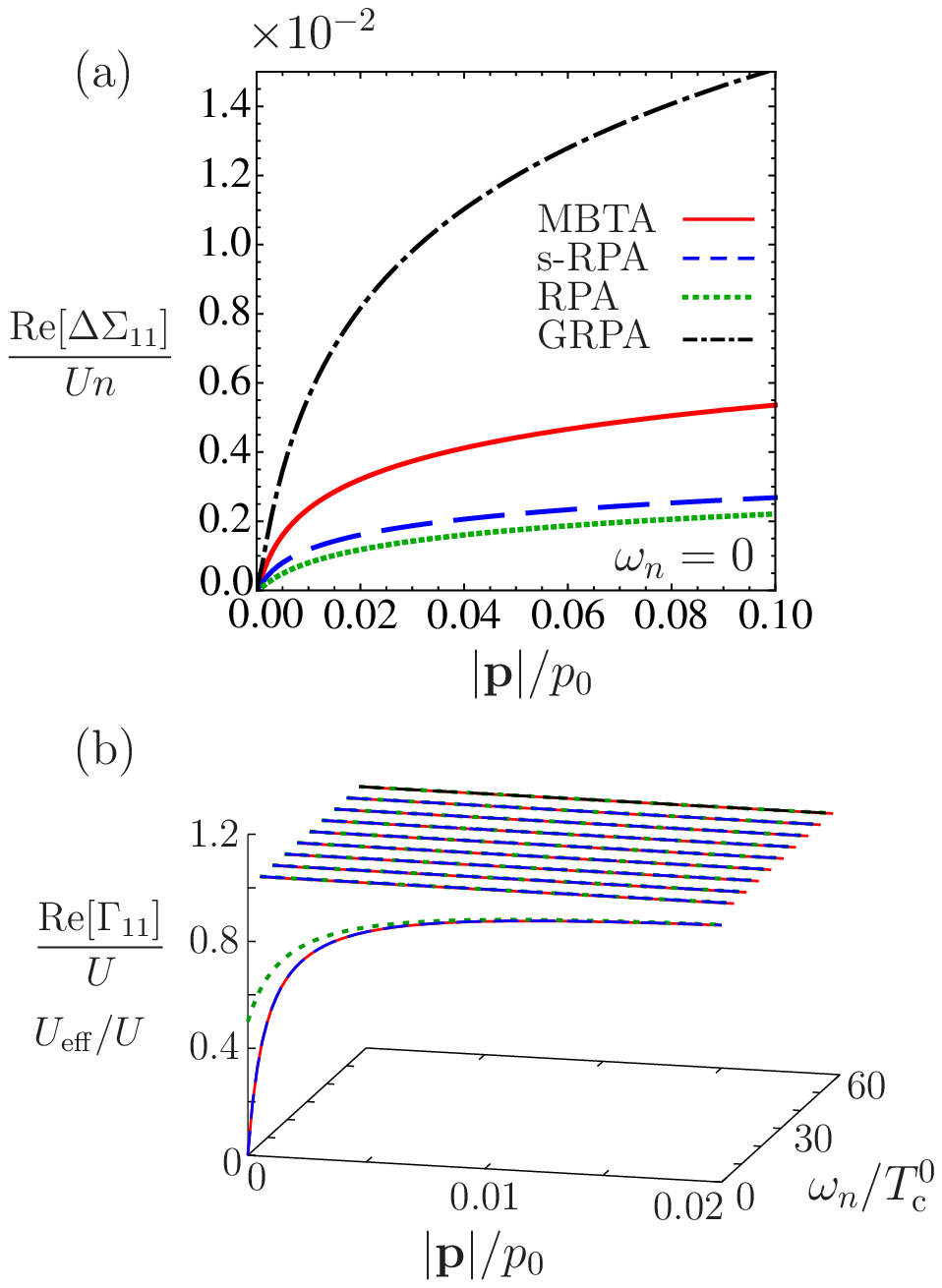}
\end{center}
\caption{(Color online) (a) ${\rm Re}[\Delta \Sigma_{11}({\bf p},0)]$, where $\Delta \Sigma_{11}({\bf p},0) \equiv \Sigma_{11}({\bf p},0)-\Sigma_{11}(0,0)$.  (b) Particle-particle scattering vertex ${\rm Re}[\Gamma_{11}]$ and the effective interaction ${\rm Re}[U_{\rm eff}]$ in the random-phase approximation (RPA) and simplified RPA (s-RPA). ${\rm Re}[U_{\rm eff}]$ in the s-RPA is almost the same as $\Gamma_{11}$ in this panel. We take $T=T_{\rm c}$, $an^{1/3} = 10^{-4}$ and $p_{\rm c} = 10 p_{0}$. 
}
\label{fig7}
\end{figure} 
\par
To understand the difference among the four results in Fig.~\ref{fig6}, it is worth noting that, when the self-energy $\Sigma_{11}$ takes a constant value, one never obtains any many-body correction to $T_{\rm c}$. In addition, the self-energy in the low-energy and low-momentum regime is a key to understanding the many-body correction to $T_{\rm c}$, because single-particle excitations in this regime are crucial for the BEC phase transition. Indeed, the increase in ${\rm Re}[\Sigma_{11}({\bf p},i\omega_n=0)]$ from the value at ${\bf p}=\omega_n=0$ is most remarkable in the GRPA, being consistent with the largest value of $c_1$ (Fig.~\ref{fig7}(a)). 
\par
As pointed out in Refs.~\cite{Shi1998,Bijlsma1996}, $\Gamma_{11}({\bf 0},0)$ vanishes at $T_{\rm c}$ (Fig.~\ref{fig7} (b)). While such a vanishing interaction is also obtained in the s-RPA, the RPA gives a finite value, $U_{\rm eff}({\bf 0},0)=U/2>0$. Apart from this difference, the effective interactions in the three approximations are almost ${\bf p}$-independent when $\omega_n\ne 0$. 
\par
From the comparison of Figs.~\ref{fig7} (b) with (a), one finds that the momentum dependence of $\Sigma_{11}$ (which is crucial for many-body corrections to $T_{\rm c}$) is dominated by the ${\bf p}$-dependence of the effective interaction in the low-energy and low-momentum region. 
\par 
\begin{figure}
\begin{center}
\includegraphics[width=10cm]{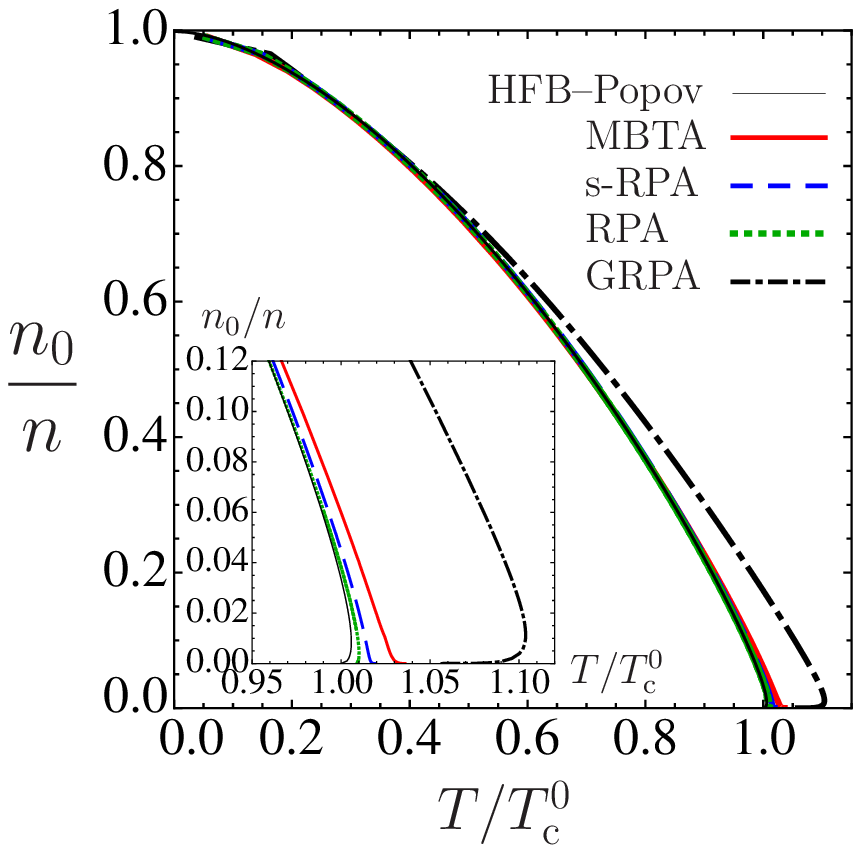}
\end{center}
\caption{(Color online) Calculated condensate fraction $n_0$ below $T_{\rm c}$. The inset shows $n_0$ magnified near $T_{\rm c}$. We set $an^{1/3} = 10^{-2}$ and $p_{\rm c} = 5 p_{0}$. The Hartree--Fock--Bogoliubov (HFB)--Popov approximation uses the interaction strength $U_0=4\pi a/m$, instead of $U$.
}
\label{fig8}
\end{figure} 
\par
\begin{figure}[t]
\begin{center}
\includegraphics[width=8cm]{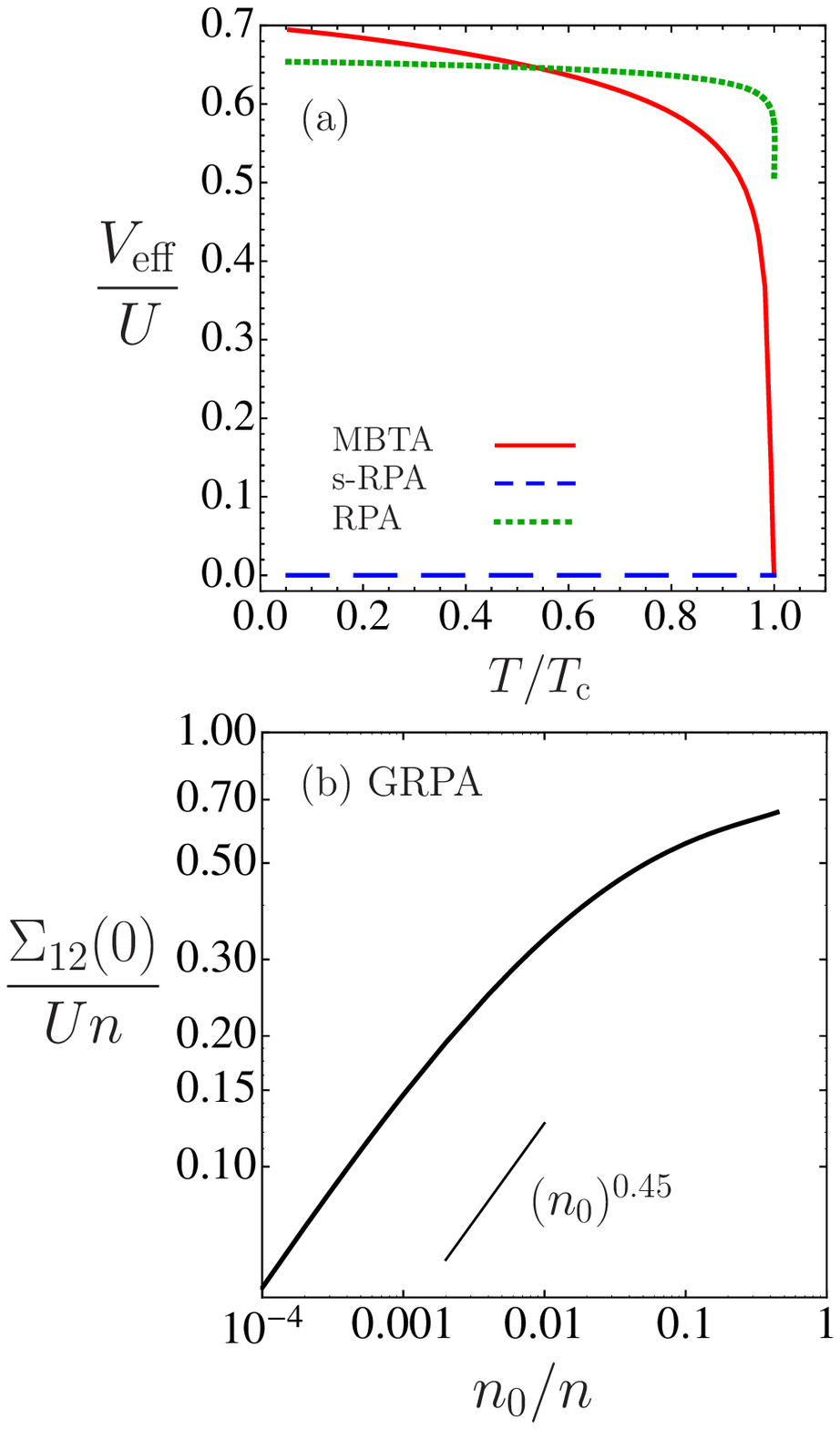}
\end{center}
\caption{(Color online) (a) Effective interaction $V_{\rm eff}(0)$, as a function of the temperature scaled by $T_{\rm c}$ in each approximation. (b) Off-diagonal self-energy $\Sigma_{12}(p=0)$, as a function of the condensation fraction $n_{0}/n$ in the generalized random-phase approximation (GRPA). We take $an^{1/3} = 10^{-2}$ and $p_{\rm c} = 5 p_{0}$. 
}
\label{fig9}
\end{figure} 
\section{Condensate fraction}
\par
While the MBTA, RPA as well as s-RPA give almost the same result as the HFB--Popov approximation (Shohno model)~\cite{Shohno1964,Popov1964A,Popov1964B,Popov1983BOOK}, the GRPA result exhibits somehow marked deviation from the mean-field result when $T\gesim 0.5T_{\rm c}^{0}$ (Fig.~\ref{fig8}). As shown in the inset, although the MBTA and s-RPA realize the expected second-order phase transition, the remaining two approximations still give the first-order phase transition, as in the HFB--Popov approximation~\cite{Reatto1969,Shi1998}. In the RPA, however, the first-order behavior is less marked than the mean-field result. The order of phase transition in these results agree with the recent work~\cite{Tsutsui2012}, which approaches $T_{\rm c}$ from the normal state.
\par 
Except for the GRPA, the off-diagonal self-energy $\Sigma_{12}$ is proportional to the effective interaction. (See Eqs. (\ref{Eq244}) and (\ref{eq28}).) In particular, at ${\bf p}=\omega_n=0$, when one simply writes the effective interactions as $V_{\rm eff}$ (which equals $\Gamma_{11}(0)$ in the MBTA, and $U_{\rm eff} (0)$ in the RPA and s-RPA), $\Sigma_{12}(0)$ in these three approximations formally has the same form as
\begin{equation}
\Sigma_{12}(0)=n_0V_{\rm eff}.
\label{eq.self120}
\end{equation}
In the mean-field theory, the first-order behavior at $T_{\rm c}$ is known to become more marked for a stronger repulsive interaction $U_0=4\pi a/m$. In this sense, the suppression of the effective interaction in the low-energy and low-momentum region shown in Figs.~\ref{fig7} (b) and \ref{fig9} (a) is favorable to the expected second-order phase transition. Indeed, both the MBTA and s-RPA (that exhibit the vanishing effective interaction at ${\bf p}=\omega_n=0$) give the second-order phase transition as shown in Fig.~\ref{fig8}. In contrast, the effective interaction at $p\to 0$ remains finite in the RPA (which gives the first-order phase transition).
\par
For the static part of the off-diagonal self-energy $\Sigma_{12}(p=0)$ in the GRPA, we find that $\Sigma_{12}(0) \propto n_{0}^{0.45}$ near $T_{\rm c}$ (Fig.~\ref{fig9} (b)), which indicates that $\Sigma_{12}(0)$ in the GRPA approaches zero near $T_{\rm c}$ more slowly than the cases of the HFB--Popov approximation and the RPA (that give $\Sigma_{12}(0)\propto n_{0}$). As a result, the first-order phase transition is more marked in the former approximation than the latter two theories, as shown in Fig.\ref{fig8}. 
\par
\section{Nepomnyashchii-Nepomnyashchii identity} 
\par
Nepomnyashchii and Nepomnyashchii proved that the off-diagonal self-energy in the low-energy and low-momentum limit exactly vanishes below $T_{\rm c}$~\cite{Nepomnyashchii1975,Nepomnyashchii1978}. This Nepomnyashchii-Nepomnyashchii (NN) identity is, however, not satisfied in the mean-field theory, where $\Sigma_{12}=n_0U>0$. In the three many-body theories (MBTA, RPA and s-RPA) giving Eq. (\ref{eq.self120}), the condition for the NN identity requires the vanishing $V_{\rm eff}$ in the BEC phase, because the condensate fraction $n_0$ is finite below $T_{\rm c}$. 
\par
Among the three theories, the s-RPA only satisfies this requirement, as shown in Fig.~\ref{fig9} (a). In the MBTA, although $V_{\rm eff}$ vanishes at $T_{\rm c}$, it becomes finite below $T_{\rm c}$, leading to the breakdown of this identity. $V_{\rm eff}$ is already finite at $T_{\rm c}$ in the RPA, so that this approximation also contradicts with this identity. The GRPA also does not satisfy the NN identity, as shown in Fig.~\ref{fig9} (b). 
\par 
While the MBTA does not satisfy the NN identity, it gives the expected second-order phase transition. In this sense, the NN identity is not necessary for the second-order phase transition to be obtained. However, as expected from the result for the s-RPA, the vanishing off-diagonal self-energy at $p=0$ itself seems favorable to the second-order phase transition. 
\par
\begin{table}[t]
\begin{center}
\caption{Aspects of four many-body approximations discussed in this paper.}    
\begin{tabular}{ccccccc}
\hline
\hline
\\[-10pt]
&& $\Delta T_{\rm c}/T_{\rm c}^{0}= c_{1} an^{1/3}$  && phase transition && NN identity\\ 
\hline
\\[-10pt]
\shortstack{GRPA} &&  $c_{1} = 6.7$ && 1st order && \xmark  \\
\shortstack{MBTA} &&  $c_{1} = 3.9$ && 2nd order && \xmark  \\
\shortstack{s-RPA } &&  $c_{1} = 2.1$ && 2nd order && \cmark \\
\shortstack{RPA}   && $c_{1} = 1.1$ && 1st order     && \xmark \\
\hline
\hline
\end{tabular} 
\end{center}
\label{tableI}
\end{table} 
\par
\section{Summary} 
To summarize, we have presented comparative studies on four kinds of many-body theories for a weakly interacting condensed Bose gas. We have treated the generalized random phase approximation (GRPA), and the many-body $T$-matrix approximation (MBTA) involving multi-scattering processes in the particle-particle scattering channel, as well as the two kinds of random phase approximations (with (RPA) and without (s-RPA) vertex corrections to each bubble diagram) describing density fluctuations. 
To treat them, we developed the 4$\times$4 matrix formalism, which includes all the possible polarization functions in four-point vertex functions.
\par 
For these approximate theories, we have examined the phase transition temperature $T_{\rm c}$, the order of phase transition (the first-order or the second-order transition), and the Nepomnyashchii-Nepomnyashchii identity, stating the vanishing off-diagonal self-energy in the low-energy and low-momentum limit. Our results are summarized in Table I, which indicates that each theory still has room for improvement. For example, while the simplified random phase approximation satisfies both the second-order phase transition and the Nepomnyashchii-Nepomnyashchii identity, the calculated $T_{\rm c}$ is found to be somehow overestimated, when the result is compared with the Monte-Carlo result. 
Since the construction of a reliable and consistent many-body theory is a crucial issue in the field of the interacting Bose gases, our results would be helpful in considering how to improve these theories for this purpose. 
\par 
Other than the above, the following issues can be raised. An interesting challenge is 
the extension of the present formalism to satisfy both the conservation law and the gapless excitation. 
The difficulty constructing the number-conserving and gapless approximation~\cite{Griffin1996,Kita20052006,Yukalov2006A,Yukalov200620082011} is known as the Hohenberg--Martin dilemma~\cite{Hohenberg1965}. 
Since we determined the chemical potential based on the Hugenholtz-Pines relation~\cite{Hugenholtz1959}, 
the excitation is gapless. 
However, the number conservation does not hold, since we used the Green's function (\ref{EqE1}) for the polarization functions, where the anomalous average is absent in the self-energy.  
\par 
Another challenging extension is to develop the two-particle irreducible (2PI) effective action approach with including many-body corrections. 
In association with the issue mentioned above, the conservation law is important for dynamics. 
The Hartree--Fock--Bogoliubov approximation~\cite{Girardeau1959} 
and others~\cite{Gardiner1997,Proukaksi2008,Cockburn2011} are theories satisfying the number-conservation. 
Among approaches for dynamics~\cite{Rey2004,Rey2005,Alexander2008,Walser1999,Walser2000,Wachter2001,Zaremba1999,Griffin2009}, the 2PI effective action approach provides a systematic method to derive the equations of motion of condensates and excitations 
with satisfying the conservation law. In fact, this formalism is a $\Phi$-derivable approximation~\cite{Hohenberg1965,Baym1962}. 
\par 
Applying this work to low-dimensional Bose gases is also interesting. 
For the system dimensionality $d = 2$, the Bose--Einstein condensation does not occur in the thermodynamic limit at non-zero temperatures, 
because of the Mermin--Wagner--Hohenberg theorem~\cite{Mermin1966,Hohenberg1967}.  
A phase transition in the thermodynamic limit in this dimensionality is the Kosterlitz--Thouless transition~\cite{Kosterlitz1973}. 
An interesting prospect is to discuss the quasicondensate in low-dimensionality using the modified Popov approximation~\cite{Andersen2002,Khawaja2002} with including many-body effects beyond the static limit. 
\par
\acknowledgements 
We thank R. Watanabe for discussions, and T. Nikuni and M. Ueda for comments. S.W. was supported by 
JSPS KAKENHI Grant Number (249416). Y.O. was supported by Grant-in-Aid for Scientific research from MEXT in Japan (25400418, 25105511, 23500056).

\appendix
\section{Polarization Functions}
\label{AppendixA}
\par 
We summarize the polarization functions used in this paper as follows: 
\begin{widetext}
\begin{align} 
\Pi_{11} (q) = & 
- 
\sum\limits_{\bf p} 
\frac{1}{2} \left [ (E_{{\bf p}+{\bf q}} - E_{\bf p}) \left ( 1 - \frac{\xi_{{\bf p}+{\bf q}}\xi_{\bf p}}{E_{{\bf p}+{\bf q}}E_{\bf p}} \right ) + i \omega_{n} \left ( \frac{\xi_{{\bf p}+{\bf q}}}{E_{{\bf p}+{\bf q}}} - \frac{\xi_{\bf p}}{E_{\bf p}}  \right ) \right ]
\frac{n_{{\bf p}+{\bf q}} - n_{\bf p}}{ \omega_{n}^{2} + (E_{{\bf p}+{\bf q}} - E_{\bf p})^{2}}
\nonumber 
\\
& \quad
-
\sum\limits_{\bf p} 
\frac{1}{2} \left [ (E_{{\bf p}+{\bf q}} + E_{\bf p}) \left ( 1 + \frac{\xi_{{\bf p}+{\bf q}}\xi_{\bf p}}{E_{{\bf p}+{\bf q}}E_{\bf p}} \right ) 
+ i \omega_{n} \left ( \frac{\xi_{{\bf p}+{\bf q}}}{E_{{\bf p}+{\bf q}}} + \frac{\xi_{\bf p}}{E_{\bf p}}  \right ) \right ]
\frac{1 + n_{{\bf p}+{\bf q}} + n_{\bf p}}{ \omega_{n}^{2} + (E_{{\bf p}+{\bf q}} + E_{\bf p})^{2}} , 
\label{gapless Hartree-Fock-BogoliubovPi11fp}
\\
\Pi_{12} (q) = & 
- \sum\limits_{\bf p} 
\frac{1}{2} \Delta \left [ \frac{\xi_{{\bf p}+{\bf q}}}{E_{{\bf p}+{\bf q}}E_{\bf p}} (E_{{\bf p}+{\bf q}} - E_{\bf p}) + \frac{ i \omega_{n} }{E_{\bf p}} \right ]
\frac{n_{{\bf p}+{\bf q}} - n_{\bf p}}{\omega_{n}^{2} + (E_{{\bf p}+{\bf q}} - E_{\bf p})^{2}}
\nonumber 
\\
& \quad
+ \sum\limits_{\bf p} 
\frac{1}{2} \Delta \left [ \frac{\xi_{{\bf p}+{\bf q}}}{E_{{\bf p}+{\bf q}}E_{\bf p}} (E_{{\bf p}+{\bf q}} + E_{\bf p}) + \frac{ i \omega_{n} }{E_{\bf p}} \right ]
\frac{1 + n_{{\bf p}+{\bf q}} + n_{\bf p}}{ \omega_{n}^{2} + (E_{{\bf p}+{\bf q}} + E_{\bf p})^{2}} , 
\label{gapless Hartree-Fock-BogoliubovPi12fp}
\\
\Pi_{14} (q) = & 
\sum\limits_{\bf p} 
\frac{1}{2} \frac{\Delta^{2}}{E_{{\bf p}+{\bf q}}E_{\bf p}}
\left [ 
(E_{{\bf p}+{\bf q}} - E_{\bf p})
\frac{n_{{\bf p}+{\bf q}} - n_{\bf p}}{ \omega_{n}^{2} + (E_{{\bf p}+{\bf q}} - E_{\bf p})^{2}}
- (E_{{\bf p}+{\bf q}} + E_{\bf p})
\frac{1 + n_{{\bf p}+{\bf q}} + n_{\bf p}}{ \omega_{n}^{2} + (E_{{\bf p}+{\bf q}} + E_{\bf p})^{2}}
\right ] , 
\label{gapless Hartree-Fock-BogoliubovPi14fp}
\\
\Pi_{22} (q) = & 
\sum\limits_{\bf p} 
\frac{1}{2} \left [ (E_{{\bf p}+{\bf q}} - E_{\bf p}) \left ( 1 + \frac{\xi_{{\bf p}+{\bf q}}\xi_{\bf p}}{E_{{\bf p}+{\bf q}}E_{\bf p}} \right ) 
+ i \omega_{n} \left ( \frac{\xi_{{\bf p}+{\bf q}}}{E_{{\bf p}+{\bf q}}} + \frac{\xi_{\bf p}}{E_{\bf p}}  \right ) \right ]
\frac{n_{{\bf p}+{\bf q}} - n_{\bf p}}{ \omega_{n}^{2} + (E_{{\bf p}+{\bf q}} - E_{\bf p})^{2}}
\nonumber 
\\
& \quad
+
\sum\limits_{\bf p} 
\frac{1}{2} \left [ (E_{{\bf p}+{\bf q}} + E_{\bf p}) \left ( 1 - \frac{\xi_{{\bf p}+{\bf q}}\xi_{\bf p}}{E_{{\bf p}+{\bf q}}E_{\bf p}} \right ) 
+ i \omega_{n} \left ( \frac{\xi_{{\bf p}+{\bf q}}}{E_{{\bf p}+{\bf q}}} - \frac{\xi_{\bf p}}{E_{\bf p}}  \right ) \right ]
\frac{1 + n_{{\bf p}+{\bf q}} + n_{\bf p}}{ \omega_{n}^{2} + (E_{{\bf p}+{\bf q}} + E_{\bf p})^{2}} , 
\label{gapless Hartree-Fock-BogoliubovPi22fp} 
\end{align}
\end{widetext}
where $\xi_{\bf p} \equiv \varepsilon_{\bf p} + \Delta$, $\Delta \equiv U n_{0}$, $E_{\bf p} \equiv \sqrt{\varepsilon_{\bf p} (\varepsilon_{\bf p} + 2 \Delta )}$, 
and $n_{\bf p}$ is the Bose distribution function $n_{\bf p} \equiv 1/ ( e^{\beta E_{\bf p}} - 1)$.  
\par
\par
\section{Vertex functions in static and zero-momentum limit}
\label{AppendixB}
We evaluate the vertex functions $\Gamma_{ij}$ in the static and low-momentum limit. Noting that $g_{11} (p)=-g_{12}(p)$ in this limit~\cite{Gavoret1964}, we find
\begin{align}
\lim_{p\to0} \Pi_{11,22,14} (p) = - \lim_{p\to0} \Pi_{12} (p). 
\label{eq1B}
\end{align} 
For the dimensionality of the system $d =3$ at $T\neq 0$, all the polarization functions show the infrared divergence 
as $\Pi_{ij} ({\bf p}, 0) \propto 1/|{\bf p}|$ for small ${\bf p}$. 
Because of these properties, the vertex functions $\Gamma_{ij}$ in the limit $p\to 0$ converge as
\begin{align}
\begin{pmatrix}
\Gamma_{11} (0)  
\\
\Gamma_{12} (0)  
\\
\Gamma_{14} (0)  
\\
\Gamma_{22} (0) 
\\
\Gamma_{23} (0)  
\end{pmatrix}
= & 
\frac{1}{2}
\begin{pmatrix}
\Gamma_{}' (0) + \Gamma_{11}' (0)  
\\
\Gamma' _{} (0)  
\\
\Gamma' _{} (0) - \Gamma_{11}'  (0) 
\\
\Gamma' _{} (0) + \Gamma_{22}'  (0) 
\\
\Gamma_{}'  (0) - \Gamma_{22}' (0) 
\end{pmatrix}.
\end{align}
Here,
\begin{align}
\Gamma_{ii}' (p) = & \frac{U}{1 - U \Pi_{ii}' (p)},  \quad (i = 1,2) 
\\
\Gamma_{}' (p) =  & \frac{U}{2 - U \Pi' (p)}, 
\end{align}
where
\begin{align}
\Pi_{ii}' (p) = & \Pi_{ii} (p) - \Pi_{14} (p),  \quad (i = 1,2) 
\\
\Pi' (p) = & 
\Pi_{11} (p) + \Pi_{22} (p) + 2 \Pi_{14} (p) + 4 \Pi_{12} (p) . 
\end{align} 
Note that $\Pi_{ii}' (p) $ and $\Pi' (p)$ converge because the infrared divergences of $\Pi_{ij}(q\to 0)$ are canceled out owing to the relation in Eq. (\ref{eq1B}). 
\par
In the same manner, the effective interaction $U_{\rm eff}(p)$, and the regular part of the density-density correlation function $\chi_{\rm R} (p)$, as well as the three-point vertices $\boldsymbol{\mathit\gamma} (p)$ and $\boldsymbol{\mathit\gamma}^{\dag} (p) $, are also found to converge in the static and zero-momentum limit, as 
\begin{align}
U_{\rm eff} (0) = & 
U \frac{2 - U \Pi' (0)}{ 3 - 2 U \Pi' (0) }, 
\\
\chi_{\rm R} (0) 
= & - \frac{1}{U} \frac{1 - U \Pi' (0) }
{ 2 - U \Pi' (0) }, 
\label{RPAPiR0}
\\
\boldsymbol{\mathit\gamma} (0) = & 
\frac{\Gamma' (0)}{U} 
\begin{pmatrix}
1 \\ 1\\ 1 \\ 1
\end{pmatrix}
- 
\begin{pmatrix}
0 \\ 1\\ 1 \\ 0
\end{pmatrix}. 
\end{align} 
Since all the vertex functions neither vanish nor diverge in the static and zero-momentum limit, $\Sigma_{0}$ also converges as
\begin{align} 
\Sigma_{011} (0) = U n_{0} \frac{A_{\Sigma 11}}{B_{\Sigma }}, \quad 
\Sigma_{012} (0) = U n_{0} \frac{A_{\Sigma 12}}{B_{\Sigma }}, 
\end{align} 
where  
\begin{align} 
A_{\Sigma 11} = & 
6 + 9 U \Pi' (0) - U^{2} [16 \Pi_{11}' (0) + 5 \Pi' (0)] \Pi' (0) 
+ 7 U^{3} \Pi_{11}' (0) \Pi'^{2} (0), 
\nonumber 
\\
A_{\Sigma 12} 
= & 
6 - U [12 \Pi_{11}' (0) - 9 \Pi' (0)] 
- U^{2} [2 \Pi_{11}' (0) + 5 \Pi' (0)] \Pi' (0) 
+ 3 U^{3} \Pi_{11}' (0) \Pi_{}'^{2} (0), 
\nonumber 
\\
B_{\Sigma}
= & 
[1 - U \Pi_{11}' (0) ] [2 - U \Pi' (0) ] [3 - 2 U \Pi' (0)].  
\nonumber
\end{align}
Since $A_{\Sigma 11,\Sigma12} $ and $B_{\Sigma }$ remain finite in the limit $p\rightarrow 0$, we find $\Sigma_{011} (0)\neq 0$ and $\Sigma_{012} (0) \neq 0$. 


\begin{thebibliography}{99}
\bibitem{NOTE} In this paper, we call diagrams that cannot be separated into two pieces by cutting a particle line or an interaction line `regular part'.
\end{thebibliography}
\end{document}